# Physical interpretation of the Planck's constant based on the Maxwell theory

**Donald C. Chang**

Macro-Science Group, Division of LIFS
Hong Kong University of Science and Technology, Clear Water Bay, Hong Kong

Email: bochang@ust.hk

**Abstract.** The discovery of the Planck's relation is generally regarded as the starting point of quantum physics. The Planck's constant $h$ is now regarded as one of the most important universal constants. The physical nature of $h$, however, has not been well understood. It was originally suggested as a fitting constant to explain the black-body radiation. Although Planck had proposed a theoretical justification of $h$, he was never satisfied with that. To solve this outstanding problem, we used the Maxwell theory to directly calculate the energy and momentum of a radiation wave packet. We found the energy of the wave packet is indeed proportional to its oscillation frequency. This allows us to derive the value of the Planck's constant. Furthermore, we showed that the emission and transmission of a photon follows the principle of all-or-none. The "strength" of the wave packet can be characterized by $\zeta$, which represents the integrated strength of the vector potential along a transverse axis. We reasoned that $\zeta$ should have a fixed cut-off value for all photons. Our results suggest that a wave packet can behave like a particle. This offers a simple explanation to the recent satellite observations that the cosmic microwave background follows closely the black-body radiation as predicted by the Planck's law.

PACS: 03.50.De; 03.65.-w; 03.65.Ta

## 1. Introduction

The birth of quantum mechanics is commonly attributed to the discovery of the Planck's relation. In order to explain black-body radiation, Planck postulated that the radiation energy is transmitted in packages ("energy quanta"). Einstein later studied the photoelectric effect and found that the energy of light absorbed by an electron is also in small "packets", which, like Planck's "energy quanta", is proportional to the light frequency $v$ [1]. This relation is now called the Planck's relation or Planck–Einstein relation:

$$E = hv. \qquad (1)$$

The constant "$h$" here is called the "Planck's constant". Subsequently, the Planck's constant was found to play a major role in many aspects of quantum physics, including the de Broglie relation $p = \hbar k$; Dirac's fundamental quantum condition: $q_r p_s - p_s q_r = i\hbar \delta_{rs}$; and Heisenberg's Uncertainty Principle $\Delta q \Delta p \sim h$ [2-4]. It has become one of the most important universal constants in physics. But up to these days, people still do not know the exact physical meaning of $h$. The Planck's constant was more like a fitting parameter; it was not derived based on first principles. In order to get a better understanding of the physical basis of the Planck's constant, this work suggests a useful approach to derive $h$ using the Maxwell theory.

Investigating the physical origin of the Planck's constant is not only important for advancing our understanding on the foundation of quantum physics, it is also relevant to the current study of cosmology. According to the Big Bang theory [5], in the early day of the universe, it is filled with energetic photons. After the universe cooled down, these primordial photons became the Cosmic Microwave Background (CMB) that we can detect today [6]. From the recent satellite measurements, the microwave detected in the CMB follows perfectly the Planck's law of black-body radiation [7]. Thus, the microwave radiation must obey $E = hv$. Does this imply that the microwave radiation is transmitted in the form of discrete particle?

It is obvious that the microwave is a wave packet of electromagnetic radiation. It cannot be regarded as a pointed object, since its wave length is quite long. Then, how to explain the fact that CMB satisfies the Planck's relation? We think the answer is that a particle does not need to be a pointed object; a wave packet can behave like a particle. This work is to show that this can be the case.

## 2. The original concept of *h* according to Planck

When it was first proposed in 1900, Planck thought that the idea of energy quantization was "a purely formal assumption ... actually I did not think much about it..." [8]. Later, he tried to justify the Planck's relation using a very complicated theoretical argument, which was hidden in cumbersome formulism of thermodynamics and statistical mechanics [9]. A more comprehensible treatment of Planck's argument was put forward by Debye in 1910 [9, 10]. In the following, we will briefly review Planck's derivation of *h* using Debye's cleaned-up version as outlined by John Slater [9].

Planck's theory was basically to treat the emitter in the black-body radiation as a linear oscillator,

$$E = T + V = \frac{p^2}{2m} + \frac{m\omega^2}{2}q^2 . \qquad (2)$$

Here, *E*, *T* and *V* represent the total energy, kinetic energy and potential energy, respectively; *q* is the generalized coordinate and *p* is the generalized momentum; *m* and *ω* are the mass and frequency of the oscillator. If one maps the energy distribution in the phase space, one will find that the contour of a constant energy is an ellipse (Figure 1a), since equation (2) can be reduced into

$$\frac{p^2}{a^2} + \frac{q^2}{b^2} = 1, \qquad (3)$$

where $a = \sqrt{2mE}$, $b = \sqrt{2E/m\omega^2}$. The area enclosed by this ellipse is known to be

$$\pi ab = \frac{2\pi E}{\omega} = \frac{E}{v} . \qquad (4)$$

Then, when the energy is increased from *E* to *E* + Δ*E*, the corresponding incremental area in the phase space is (see Figure 1a)

$$\Delta A = \Delta E/v. \qquad (5)$$

Planck then made two formal assumptions:

*(i) The energy distribution of the emitter in reality does not follow a smooth curve as described in equation (2). Instead, it is a step-wise function with a constant jump of magnitude ΔE (Figure 1b).*
*(ii) This jump is because nature somehow partitions the phase space of the emitter in constant incremental areas (ΔA), which can be called "h" (i.e., ΔA = h). According to equation (5), this assumption implies that*

$$\Delta E = hv. \qquad (1A)$$

With the above assumption, one can calculate the average energy of a black-body radiation emitter oscillating at frequency *v* using the Boltzmann distribution, that is,

$$\text{Average Energy} = \frac{0 + hv\exp(-hv/kT) + 2hv\exp(-2hv/kT) + 3hv\exp(-3hv/kT) + ...}{1 + \exp(-hv/kT) + \exp(-2hv/kT) + \exp(-3hv/kT) + ...} \qquad (6)$$

For simplicity, let us denote $\exp(-hv/kT) \equiv x,$

$$\text{Average Energy} = \frac{h\nu x(1+2x+3x^2+...)}{1+x+x^2+x^3+...}.$$

We know $1+x+x^2+x^3+...= \frac{1}{1-x}$, and $1+2x+3x^2+... = \frac{d}{dx}(1+x+x^2+x^3+...)$.

Therefore,

$$\text{Average Energy} = \frac{h\nu x}{1-x} = \frac{h\nu \exp(-h\nu/kT)}{1-\exp(-h\nu/kT)} = \frac{h\nu}{\exp(h\nu/kT)-1}. \qquad (7)$$

Since we know the number of energy states per unit volume and per unit frequency range in the radiation field is $8\pi\nu^2/c^3$, the energy distribution in the radiator then is [9]

$$u_\nu = \frac{8\pi h\nu^3}{c^3} \frac{1}{\exp(h\nu/kT)-1}. \qquad (8)$$

This equation is now known as the "Planck's law". It fitted the experimental data of black-body radiation very well [11].

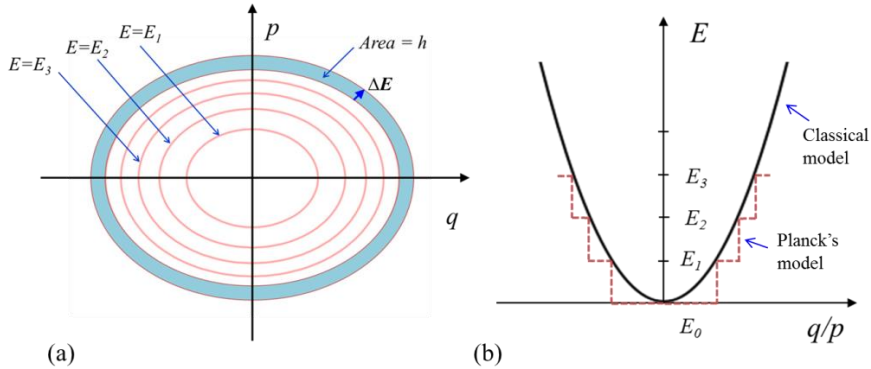

*Figure 1. **Modeling the emitter in black-body radiation as a linear oscillator.** (a) Energy distribution in phase space. The contour of equi-energy is an ellipse. (b) Planck's step-wise energy distribution model for emitter in black-body radiation. The smooth curve was assumed to become a step-wise function with a constant jump of magnitude ΔE (dotted lines).*

Although the Planck's law was a great success, Planck was not satisfied with the physical meaning of *h* as he derived it. Particularly, he knew that the assumption of partitioning the phase space (of the oscillator) in equal incremental area is somewhat arbitrary. Planck spent subsequent years trying to justify his theory on better physical grounds but was not successful [9].

## 3. Determination of energy and momentum carried by a photon

In this paper, we try to uncover the physical meaning of *h* by treating the photon as a wave packet of electro-magnetic radiation and directly calculating the total energy and momentum contained within the wave packet. More explicitly, the approach of this study is based on the following steps:

- We will regard the photon as a wave packet which is made up of an oscillating electro-magnetic field.
- The energy (*E*) and momentum (*p*) of the electro-magnetic field contained within the wave packet can be calculated based on the Maxwell theory.
- We will examine whether *E* is proportional to the oscillating frequency *v*. If yes, the proportional constant will be identified as the Planck's constant.

## 3.1. Energy density of an electro-magnetic field

The energy density of an electro-magnetic field is known to be [12]

$$U = \frac{1}{2}\left(\varepsilon \mathbf{E}^2 + \frac{1}{\mu}\mathbf{B}^2\right), \quad (9)$$

where $\varepsilon$ and $\mu$ are the dielectric permittivity and magnetic permeability of the vacuum, $\mathbf{E}$ and $\mathbf{B}$ are electric field and magnetic induction, respectively. According to the Maxwell's theory, $\mathbf{E}$ and $\mathbf{B}$ can be derived from the scalar potential $\Phi$ and the vector potential $\mathbf{A}$:

$$\begin{cases} \mathbf{B} = \nabla \times \mathbf{A} & (10A) \\ \mathbf{E} = -\nabla\Phi - \dfrac{\partial \mathbf{A}}{\partial t}. & (10B) \end{cases}$$

In electro-magnetic radiation, the vector potential $\mathbf{A}$ obeys the wave equation

$$\nabla^2 \mathbf{A} - \frac{1}{c^2}\frac{\partial^2 \mathbf{A}}{\partial t^2} = 0. \quad (11)$$

In order to calculate the energy density of an electro-magnetic wave, let us choose a simple system in which the wave is traveling along the z-axis and the vector potential is along the x-axis (see Figure 2), i.e., $\mathbf{A} = A_x \hat{x}$, and

$$A_x = A_0 e^{i(kz-\omega t)}. \quad (12)$$

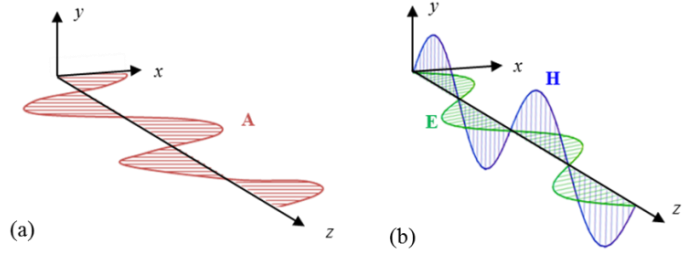

*Figure 2. A simplified model for the propagation of electro-magnetic radiation. (a) Vector potential A oscillates along the x-axis and the wave is traveling along the z-axis. (b) The electric field E oscillates in x direction; the magnetic field H is perpendicular to E and oscillates in y direction.*

Since there is no embedded charge within the vacuum, $\Phi = 0$. Thus, equations (10A) and (10B) becomes

$$\begin{cases} \mathbf{B} = \left(\dfrac{\partial A_z}{\partial y} - \dfrac{\partial A_y}{\partial z}\right)\hat{x} + \left(\dfrac{\partial A_x}{\partial z} - \dfrac{\partial A_z}{\partial x}\right)\hat{y} + \left(\dfrac{\partial A_y}{\partial x} - \dfrac{\partial A_x}{\partial y}\right)\hat{z} = \dfrac{\partial A_x}{\partial z}\hat{y} & (13A) \\ \mathbf{E} = -\dfrac{\partial \mathbf{A}}{\partial t} = -\dfrac{\partial A_x}{\partial t}\hat{x}. & (13B) \end{cases}$$

Substituting equations (13A) and (13B) into equation (9), we have

$$U = \frac{1}{2}\left[\varepsilon\left|\frac{\partial A_x}{\partial t}\right|^2 + \frac{1}{\mu}\left|\frac{\partial A_x}{\partial z}\right|^2\right]. \quad (14)$$

This relation suggests that $A_x$ is playing the role of the "field parameter" in wave propagation. This point can be easily seen by comparing equation (14) with the energy density equation in a one-dimensional stretched string (Figure 3), which is

$$U = \frac{1}{2}\rho\left(\frac{\partial \phi}{\partial t}\right)^2 + \frac{1}{2}F_1\left(\frac{\partial \phi}{\partial z}\right)^2. \qquad (15)$$

Here $\rho$ is the mass density of the string, and $F_1$ is the tension of the string [13]. One can immediately see that, in the electro-magnetic system, $A_x$ appears to play the role of a propagating field, just like the displacement $\phi$ in the stretched string.

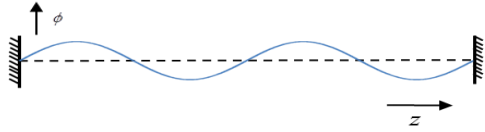

*Figure 3. **Wave propagation in a 1-D stretched string**. The wave is traveling along the z axis and $\phi$ is the local displacement of the string.*

Recall that the speed of light $c = \omega/k = \sqrt{1/\varepsilon\mu}$, one can directly calculate the energy density of the electro-magnetic system from equations (12) and (14),

$$U = \frac{1}{2}\left[\varepsilon(\omega A_0)^2 + \frac{1}{\mu}(kA_0)^2\right] = \varepsilon\omega^2 A_0^2. \qquad (16)$$

### 3.2. Total energy contained in a wave packet representing a photon

To find the physical meaning of the Planck's constant, we need to calculate the total energy $\langle U \rangle$ contained within the wave packet representing one single photon. This energy can be obtained directly by integrating the energy density described in equation (16) over the entire volume of the wave packet:

$$\langle U \rangle = \iiint_v U(x, y, z)\,dxdydz. \qquad (17)$$

In order to carry out this integration, one must know the structure of a photon. In the literature, a photon is usually described by equation (12). This, however, is not strictly correct since it represents a continuous wave; which spreads over the entire space and time. The photon should have a limited size along its trajectory (z-axis) and in the transverse plane (xy plane). It should be a wave packet, which is constructed by superposition of multiple wave components. Figure 4 shows three basic types of traveling waves: (a) *A continuous wave*. The wave frequency is a fixed constant. (b) *A wave packet with limited spread on the space and time dimensions* ($\Delta\omega$ is very small). (c) *A wave packet with a narrow spread over space and time* ($\Delta\omega$ is very large). What does a photon look like? Since a coherent light (such as a laser) has very narrow linewidth, the wave packet of a photon must be similar to that shown in Figure 4b. Such a wave function can be written as

$$A_{wp} = A_0(x, y, z, t)e^{i(kz-\omega t)}, \qquad (18)$$

where $A_0(x, y, z, t)$ describes the envelope of the wave packet. Since the photon travels in a straight line along the z-axis, we can assume that it has axis symmetry, i.e.,

$$A_0(x, y, z, t) = A_T(r,\theta)A_L(z-ct), \qquad (19)$$

where $A_T$ and $A_L$ are the transverse and longitudinal component, respectively. Equation (18) then becomes

$$A_{wp} = A_T(r,\theta)\underbrace{A_L(z-ct)e^{i(kz-\omega t)}}_{A_{path}}. \qquad (20)$$

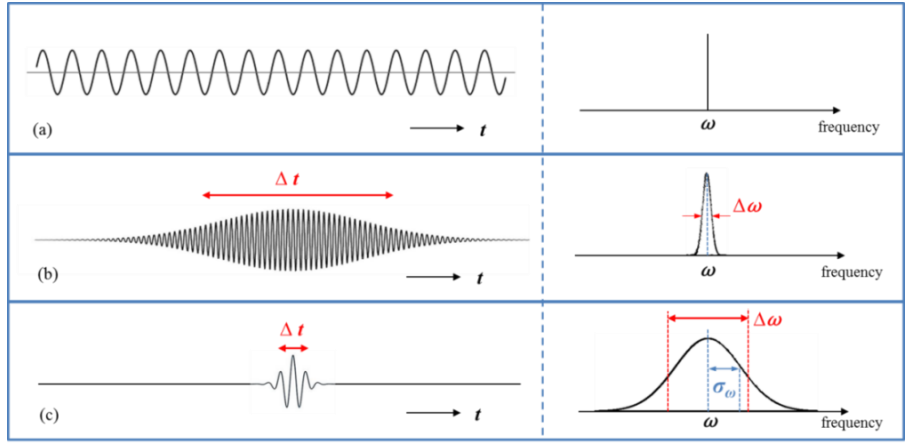

*Figure 4. Three basic types of traveling wave. The left-side is plotted in time domain; the right-side is plotted in frequency domain. (a) A plane wave; its frequency ω is a constant. (b) A wave packet with narrow linewidth; $\Delta\omega$ is very small in comparison to its average frequency ω. The half-width of the wave packet in the time domain is $\Delta t$. (c) A wave packet with large linewidth. Here, we also show the standard deviation $\sigma_\omega$ of the Gaussian function.*

$A_{path}$ can be constructed by superposition of plane waves with frequency slightly different from the central frequency ($\omega$), i.e.,

$$A_{path}(z,t) \approx \int_0^\infty g(\omega')e^{i(kz-\omega't)}d\omega', \qquad (21)$$

where $g(\omega')$ is the frequency distribution function, which can be assumed to follow a Gaussian distribution with a standard deviation $\sigma_\omega$, i.e.,

$$g(\omega') \sim e^{-(\omega'-\omega)^2/2(\sigma_\omega)^2}. \qquad (22)$$

It is well known that the Fourier transform of which will also give a Gaussian distribution in the time domain. That is,

$$A_L(z-ct) \sim e^{-(z-ct)^2/2c^2(\sigma_t)^2}, \qquad (23)$$

and

$$\sigma_\omega \cdot \sigma_t = 1. \qquad (24)$$

Once we know the envelope function in the time domain, we can easily obtain the envelope function in the spatial domain along the $z$-axis,

$$A_L(z-ct) \sim e^{-(z-ct)^2/2(\sigma_z)^2}, \qquad (25)$$

where $\sigma_z = c \cdot \sigma_t$. Based on equation (16) and equation (17), we can now calculate the total energy of the wave packet:

$$\langle U \rangle = \iiint_v (\varepsilon \omega^2 A_0^2) \, dx\, dy\, dz$$

$$= \varepsilon \omega^2 \underbrace{\int_0^\infty \int_0^{2\pi} A_T^2(r,\theta) \, r\, d\theta\, dr}_{\rho_T} \underbrace{\int_{-\infty}^\infty A_L^2(z-ct) \, dz}_{\rho_L}. \tag{26}$$

$\rho_T$ and $\rho_L$ in equation (26) can be calculated separately:

$$\rho_L = \int_{-\infty}^\infty e^{-(z-ct)^2/(\sigma_z)^2} \, dz = \sqrt{\pi}\,(\sigma_z). \tag{27}$$

Recall that $\sigma_z = c\,\sigma_t$ and $\sigma_\omega \cdot \sigma_t = 1$,
$$\sigma_z = c \cdot \sigma_t = c\,\frac{1}{\sigma_\omega}. \tag{28}$$

$\sigma_\omega$ is known to be related to the linewidth (or half-width, $\Delta\omega$) of the photon,

$$\Delta\omega = 2\sqrt{2\ln 2}\,\sigma_\omega = 2.355\,\sigma_\omega. \tag{29}$$

In most transmitting media, the linewidth of a wave is proportional to the frequency $\omega$. This ratio is defined as the "$Q$ factor",

$$Q = \frac{\omega}{\Delta\omega}. \tag{30}$$

The value of the $Q$ factor is determined by the properties of the transmitting medium. Combining equations (28), (29) and (30), we have

$$\sigma_z = c\,\frac{2.355}{\Delta\omega} = \frac{2.355\,cQ}{\omega}.$$

Substituting this into equation (27), and recall $c = \sqrt{1/\varepsilon\mu}$, we have

$$\langle U \rangle = 2.355\sqrt{\pi}\,\varepsilon\omega^2 \,\frac{cQ}{\omega}\,\rho_T = 2.355\,Q\sqrt{\frac{\pi\varepsilon}{\mu}}\,\rho_T\,\omega. \tag{31}$$

Next, we need to calculate the value of $\rho_T$. Its value can be easily determined if one knows the functional form of $A_T$. As we had discussed earlier, the size of the wave packet in the transverse plane cannot be infinite. The simplest way to model $A_T$ is to assume that it has a constant value up to a cut-off radius ($r_0$). $A_T$ then vanishes when $r > r_0$. A more reasonable model, however, is to assume that $A_T$ follows a bell-shape Gaussian distribution (Figure 5), i.e.,

$$A_T(r,\theta) = a e^{-r^2/2\sigma^2}, \tag{32}$$

where $a$ is the amplitude of the envelope function. From equations (26) and (32),

$$\rho_T = \int_0^\infty \int_0^{2\pi} a^2 e^{-r^2/\sigma^2} \, r\, d\theta\, dr = 2\pi a^2 \int_0^\infty e^{-r^2/\sigma^2} \, r\, dr = \pi \sigma^2 a^2. \tag{33}$$

This is closely related to the area of integrating the transverse component $A_T$ along the $x$-axis, which we can call it "$\varsigma$" (see Figure 5b),

$$\zeta \equiv \int_{-\infty}^{\infty} A_T\, dx = \int_{-\infty}^{\infty} ae^{-x^2/2\sigma^2}\, dx = \sqrt{2\pi}\, a\sigma. \qquad (34)$$

Combining equations (33) and (34), we have
$$\rho_T = \frac{1}{2}\zeta^2. \qquad (35)$$

Substituting this result into equation (31), and recall $\omega = 2\pi v$, we get

$$\langle U \rangle = 2.355 Q \sqrt{\frac{\varepsilon\pi}{\mu}} \frac{1}{2}\zeta^2 \omega = \left(13.113 Q \sqrt{\frac{\varepsilon}{\mu}}\zeta^2\right) v. \qquad (36)$$

Since $\langle U \rangle$ represents the total electro-magnetic energy of a single photon, equation (36) is identical to the Planck's relation $E = hv$, and the Planck's constant $h$ can be identified as

$$h = 13.113 Q \sqrt{\frac{\varepsilon}{\mu}}\zeta^2 . \qquad (37)$$

We will show later that $\zeta^2$ has a clear meaning in quantum physics and should have a fixed cut-off value for a photon (see **Section 4**).

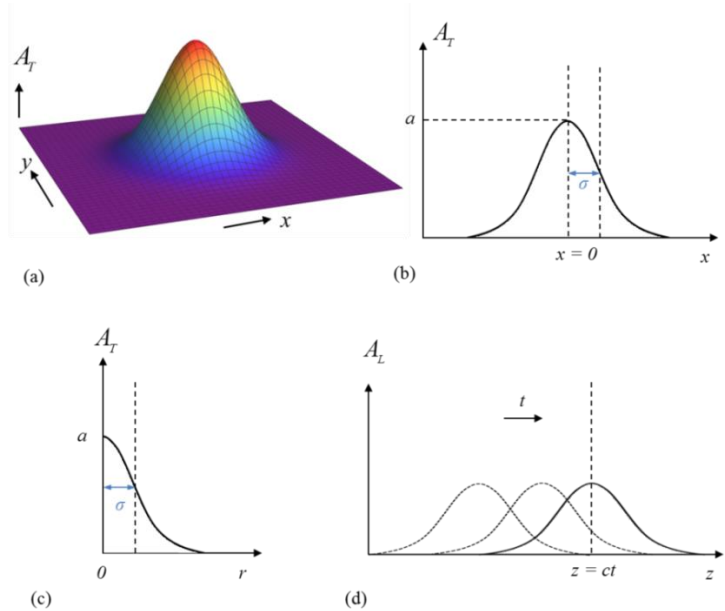

*Figure 5. A Gaussian distribution model of an electro-magnetic wave packet. (a) A 2-dimensional plot of the transverse component of the envelope function, $A_T$. (b) The cross-section plot of $A_T$ along the x-axis; $\sigma$ is the standard deviation; a is the wave amplitude at the peak. (c) Variation of $A_T$ along the radius dimension. (d) A plot of the longitudinal component of the envelope function, $A_L$, along the z-axis. The wave packet moves at the speed c.*

### 3.3. Total momentum contained in a wave packet: Justification of the de Broglie relation

By treating the photon as a wave packet of electro-magnetic radiation, we can also calculate the total momentum contained within the wave packet. It is well known that the energy flow of the electro-magnetic field can be described by the Poynting vector **S**: [12]

$$\mathbf{S} = \varepsilon c^2 \mathbf{E} \times \mathbf{B}. \qquad (38)$$

Since it can be shown that in a radiation system, $|\mathbf{E}| = c|\mathbf{B}|$, then,

$$\mathbf{S} = \varepsilon c^2 \mathbf{E} \times \mathbf{B} = \varepsilon c |\mathbf{E}|^2 \hat{\mathbf{z}}. \tag{39}$$

From equation (16), we know $\varepsilon |\mathbf{E}|^2 = \varepsilon \omega^2 A_0^2 = U$, equation (39) becomes

$$\mathbf{S} = cU \hat{\mathbf{z}}. \tag{39A}$$

The total energy flux of a wave packet of electro-magnetic radiation then is

$$\langle \mathbf{S} \rangle = \iiint_v \mathbf{S}\, dxdydz = \iiint_v (cU)\, dxdydz\, \hat{\mathbf{z}}$$

$$= c \langle U \rangle \hat{\mathbf{z}}. \tag{40}$$

For an electro-magnetic wave, the momentum density ($\mathbf{g}$) is known to be related to the Poynting vector $\mathbf{S}$ by [14]

$$\mathbf{g} = \frac{1}{c^2} \mathbf{S}. \tag{41}$$

Thus, the total momentum of a wave packet is

$$\langle \mathbf{g} \rangle = \iiint_v \mathbf{g}\, dxdydz = \frac{1}{c^2} \iiint_v \mathbf{S}\, dxdydz = \frac{1}{c^2} \langle \mathbf{S} \rangle. \tag{42}$$

Substituting equation (40) into equation (42), and using equation (36), we have

$$\langle \mathbf{g} \rangle = \frac{1}{c} \langle U \rangle \hat{\mathbf{z}} = 13.113 Q \sqrt{\frac{\varepsilon}{\mu}} \zeta^2 \frac{v}{c} \hat{\mathbf{z}}. \tag{43}$$

Previously, we have already identified the value of $h$ from equation (37). Recall that the wave vector $k = 2\pi/\lambda = 2\pi v/c$, equation (43) becomes

$$\langle \mathbf{g} \rangle = h \frac{v}{c} \hat{\mathbf{z}} = \frac{h}{2\pi} k \hat{\mathbf{z}} = \hbar k \hat{\mathbf{z}}. \tag{44}$$

Equation (44) shows that the total momentum of a photon is proportional to its wave vector $\mathbf{k}$. Equation (44) is identical to the de Broglie relation [2],

$$\mathbf{p} = \hbar \mathbf{k}. \tag{45}$$

Therefore, the Planck's constant derived by us not only satisfies the Planck's relation, it also satisfies the de Broglie relation.

## 4. Discussions

### *4.1. Physical meaning of h being a constant: The principle of all-or-none*

In the foregoing sections, we demonstrated that one can directly calculate the energy of a photon based on the Maxwell's theory. Based on this result, the Planck's constant is given by

$$h = 13.113 Q \sqrt{\frac{\varepsilon}{\mu}} \zeta^2. \tag{37}$$

Apparently, the Planck's constant is dependent on the physical properties of the vacuum, e.g., the dielectric permittivity $\varepsilon$ and magnetic permeability $\mu$. The quality factor $Q$ is also a property of the vacuum, since it is dependent on the transmitting medium. At this point, we do not know enough about the detailed properties of the vacuum to directly calculate $Q$. But the value of $Q$ can be determined by experiment. One can use an optical device to directly measure the linewidth of a photon with known frequency. In the literature, there were already some hints about the value of $Q$. For example, it was reported that a solid-state dye laser (at 590 nm) could have a linewidth around 350 MHz [15]. This suggests that the $Q$ factor is about $1.45 \times 10^6$. Our work may motivate more accurate measurement of $Q$ in the future.

The remaining problem is to consider whether $\zeta^2$ can be regarded as a constant and what does that mean. From equation (34), $\zeta$ is defined as the integrated area of the vector potential at the center of the wave packet. The requirement for $\zeta$ being a constant means that: *Regardless of the oscillating frequency, in order to generate a sustainable oscillating wave, nature requires a critical amount of disturbance in the electro-magnetic field.*

This situation is very similar to the generation of a nerve impulse. We know that a neuron can transmit a signal to its downstream target along its nerve fiber (called "axon"). This signal is called an "action potential" [16]. It is well known that the generation of action potential has the property of "**all-or-none**". That means, when the stimulus to the axon is below a threshold, no action potential can be generated. But when the stimulus is higher than the threshold, a full size action potential will be generated. This action potential will propagate along the axon with constant amplitude (about 100 mV) [17]. In another word, one cannot generate an action potential with arbitrarily small amplitude. And, no matter how large is the stimulus; one cannot generate an action potential which is much larger than 100 mV. That is why people called it "all-or-none".

As it turns out, this principle of all-or-none is applied in multiple aspects of nature. Not only the transmission of a nerve impulse is all-or-none, the transmission of the electro-magnetic radiation is also all-or-none. The radiation energy is apparently transmitted in small packets (photon), each of which has a limited size and is not sub-dividable. Either one can generate a full size photon, or generate no photon at all. In another word, if the energy of the electro-magnetic field is smaller than a critical value, it will not be able to trigger a transmissible excitation wave traveling as a wave packet. Instead, the energy will just dissipate in the surrounding.

This requirement of "all-or-none" means that $\zeta$ should have a fixed cut-off value; it cannot be arbitrarily small. Thus, although the size of the wave packet is not fixed, the total amount of disturbance in the electromagnetic field (as measured by $\zeta$) is fixed (see Figure 6). If the diameter of the wave packet is very small, the oscillation amplitude of the electro-magnetic field within the wave packet must be large enough to make $\zeta$ reach the threshold value. Alternatively, if the oscillation amplitude of the wave packet is small, the size of the wave packet must be large enough to compensate it so that the integrated area reaches the threshold value.

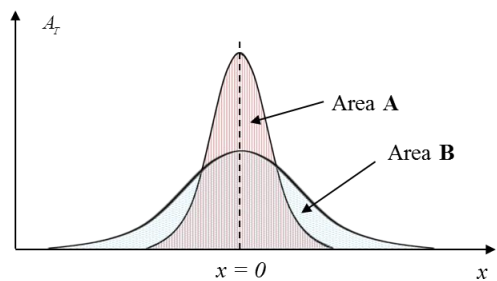

*Figure 6. The integrated area of the transverse component $A_T$ along the x-axis (denoted by "$\zeta$") is a constant during wave propagation*. **A** *and* **B** *represent the transverse components of two different wave packets with different oscillation amplitudes (a) and widths (σ). Nature requires that their integrated areas ($\sqrt{2\pi}\, a\sigma$) to be the same, i.e., Area A = Area B.*

Besides the above considerations, there is another reason suggesting that $\zeta^2$ should be a constant. Recall that $\zeta^2$ was defined by

$$\rho_T = \int_0^\infty \int_0^{2\pi} |A_T(r,\theta)|^2 \, r d\theta dr = \frac{1}{2}\zeta^2. \tag{35}$$

As we pointed out earlier in Section 3.1, the vector potential (**A**) in an electro-magnetic radiation system is equivalent to the amplitude of an oscillation wave in a one-dimensional vibrating string. Thus, if one wants to write down the wave function of a photon ($\phi$), one can guess that $\phi$ must be related to **A** (with a normalizing factor). Since the absolute square of the wave function $|\phi|^2$ is usually interpreted as the probability of finding the particle, this suggests that the absolute square of the vector potential of the wave packet, $|A_{wp}|^2$, is proportional to the probability of finding the photon. Recall that $|A_{wp}|^2 = |A_0|^2 = |A_T|^2$ at the center of the wave packet, where $z - ct = 0$, one can interpret

$$\rho_T = \int_0^\infty \int_0^{2\pi} |A_T(r,\theta)|^2 \, r d\theta dr \sim \int_0^\infty \int_0^{2\pi} |\phi|^2 \, r d\theta dr$$

as a measure of "the total probability of finding the photon at the center of the wave packet". So long as the wave packet represents a single photon, this probability should remain constant as the wave packet travels along the trajectory of the photon. Thus, the requirement of $\zeta^2$ being a constant essentially means that, in an optical measurement, *the probability of finding the photon at the center of the wave packet is always the same.*

*4.2. Reconciliation with Planck's original model*

Next, let us examine if our wave packet model is compatible with Planck's original thinking. In the original work of Planck, his proposal of $\Delta E = h\nu$ in black-body radiation was based on two assumptions: (a) The emitter can be modeled as a linear oscillator; (b) The energy distribution of this oscillator somehow exhibits a step-wise pattern (blue dashed lines on the left image in Figure 7). There were two problems with this model:

1. In order to show directly that the energy distribution of black-body radiation is quantized; one has to calculate the energy distribution of the radiation wave, not just the emitter.

2. Planck's hypothesis that the energy distribution in the linear oscillator is step-wise is an arbitrary assumption. It needs to be justified.

Now, with the findings of this paper, we may be able to explain why Planck's model could give the correct result. In black-body radiation, the emitter gives out radiation waves which will be absorbed by the detector. If one accepts that the radiation wave is transmitted in wave packets, our calculation shows that the group energy of each wave packet is proportional to its oscillation frequency, i.e.,

$$\langle U \rangle = 13.113 Q \sqrt{\frac{\varepsilon}{\mu}} \zeta^2 \nu = h\nu. \tag{36}$$

In Planck's model, the energy distribution of an emitter was modeled as a linear oscillator. It has a parabolic shape (see Figure 7). The emission of a wave packet depends on the energy transitions from a higher energy state to a lower energy state of the emitter. When the emitter's energy state is less than $h\nu$ above the ground state, it is incapable of giving off a photon at frequency $\nu$. That means that if one tries to detect the radiation signal from the emitter, one will fail to detect any photon emission (at frequency $\nu$). This situation changes only when the energy level of the emitter reaches $h\nu$ or above. Thus, if one studies the black-body radiation at a single frequency $\nu$, there is no transmission of radiation energy from the emitter to the detector until the emitter reaches $E_1$, where $E_1 - E_0 = \Delta E = h\nu$. Similarly, the emitter can be detected to emit two photons when the emitter reaches $E_2$, where $E_2 - E_1 = \Delta E = h\nu$. This explains why the emission of photon is a step-wise process when the energy of the emitter increases. (For more details, see Figure 7.)

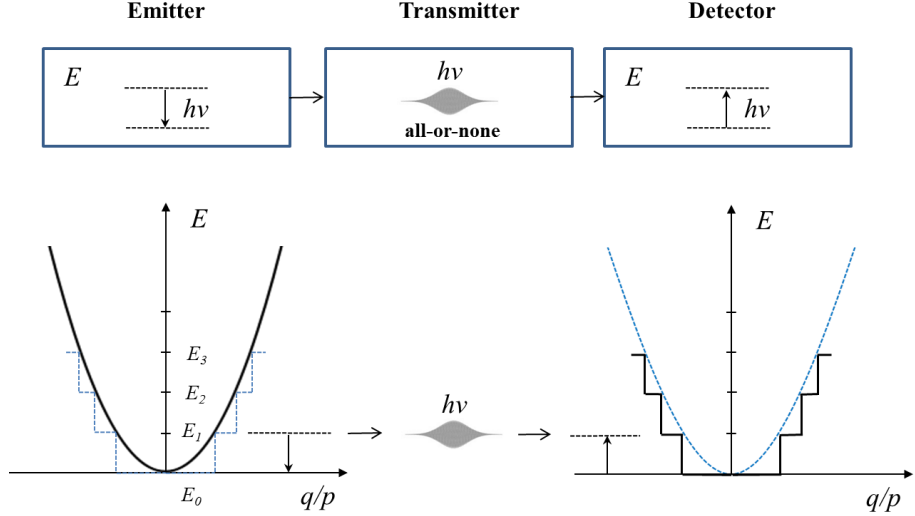

*Figure 7. Explanation of why Planck's original model worked. The black-body radiation involves three steps: emission, transmission and absorption. The emitter gives off radiation energy in individual wave packets. We showed that each wave packet contains an amount of energy hv. The emission of the wave packet is all-or-none. If the energy transition within the emitter is smaller than hv, no radiation wave could be emitted. When the energy transition reaches hv, it gives off one photon. Thus, even though the energy distribution of the emitter is continuous (solid line in the left image), its emission of photon follows a step-wise pattern (dashed lines in the left image). When a photon is emitted, it travels as a wave packet in an all-or-none fashion. Finally, it is totally absorbed by a detector. Hence, from the detector's point of view, the energy distribution appears to follow a step-wise pattern (solid line in the right image).*

Thus, even though Planck did not calculate the energy content of a radiation wave packet, he could still correctly explain the black-body radiation by using a step-wise energy distribution model. All he needed was to assume that the energy released from the emitters is in packets, which he called "energy quanta".

### 4.3. Implications on the physical meaning of Heisenberg's Uncertainty Principle

Once we can derive the Planck's relation and de Broglie's relation based on the wave packet model, Heisenberg's Uncertainty Principle can be easily explained. If one accepts that a photon is a wave packet of oscillating electro-magnetic field, which follows a Gaussian distribution along the particle trajectory, the half-width of the wave packet in the time domain can be directly determined from the linewidth of the radiation wave. From the condition of Fourier transform, we know

$$\sigma_\omega \cdot \sigma_t = 1, \quad (24)$$

where $\sigma_\omega$ and $\sigma_t$ are the standard deviations in the frequency domain and the time domain. Since we know the half-width of the wave packet is $\Delta t = 2.355\, \sigma_t$ and the linewidth of the oscillation frequency is $\Delta\omega = 2.355\, \sigma_\omega$, Equation (24) implies $\Delta\omega \cdot \Delta t = (2.355)^2$. From the Planck's relation, $E = \hbar\omega$, we have

$$\Delta E \cdot \Delta t = \hbar \Delta\omega \cdot \Delta t = \frac{h}{2\pi}(2.355)^2 = 0.8827\, h. \quad (46)$$

This suggests that the product of linewidths in the energy and time domains for a single photon is very close to $h$. Such a result agrees with Heisenberg's conjecture that

$$\Delta E \cdot \Delta t \approx h. \tag{47}$$

Thus, Heisenberg's Uncertainty Principle can be interpreted as a direct result of the fact that a photon is a wave packet which follows a Gaussian distribution.

Using this wave packet model, we can also easily obtain the Uncertainty Principle between $\Delta p$ and $\Delta z$. Recall from the de Broglie relation, $\Delta p = \hbar \Delta k = \hbar(2.355\sigma_k)$. The half-width of the wave packet is $\Delta z = 2.355\,\sigma_z$. From the conditions of the Fourier transform,

$$\sigma_k \cdot \sigma_z = 1. \tag{48}$$

Then, the above relations give

$$\Delta p \cdot \Delta z = \hbar \Delta k \cdot \Delta z = \frac{h}{2\pi}(2.355)^2 \sigma_k\, \sigma_z = 0.8827\, h. \tag{49}$$

This agrees with the conjecture of Heisenberg that $\Delta p \cdot \Delta z \approx h$.

### *4.4. Other discussions*

This work represents an approach to use a classical theory to explain the physical basis of $h$. What we have done so far was to derive the Planck's relation by calculating directly the energy contained within a wave packet based on the Maxwell's theory. Using such an approach, we can also derive the de Broglie relation and Heisenberg's Uncertainty Principle. These derivations were straight forward and only based on a simple assumption that the photon is a wave packet of electro-magnetic radiation.

Since the Planck's constant is a fundamental physical constant, it is involved in many areas of studies, including the foundation of quantum mechanics [18, 19], quantum field theory [20, 21], the study of chaos [22] and tunneling [23], etc. There have been many previous attempts to explain the physical basis of the Planck's constant [24-28]. For example, Galgani and Scott tried to use a classical mechanical model of a one-dimension particle chain to explain the Planck's relation [24]. They assumed a Lennard-Jones interaction between the nearest-neighboring particles and solved the classical equations of motion numerically. For a broad class of initial conditions, Planck-like distributions were obtained for the time averages of the energies of the normal modes. They reported that the action constant entering such a distribution is in the same order of magnitude as the Planck's constant [24].

In another study, Ross proposed a possible way of building the Planck's constant into the structure of space-time [26]. This was done by assuming that the torsional defect that intrinsic spin produces in the geometry is a multiple of the Planck length. He showed that, by using a simple geometrical assumption, it could lead to quantization of angular momentum. He thought that such an approach could be considered as a first step to derive the value of $\hbar$.

As an extension of Ross's study, Duan *et al.* [28] proposed a new geometrization of Planck's constant in terms of vierbein theory [29]. They thought that $h$ is also connected with the defects of space-time. A set of invariances including the *U(1)*-like gauge transformations was used in this geometrization. Using the gauge-potential decompositions, the quantity introduced in this geometrization to describe the defects would be quantized. Such an approach would allow one to relate the Planck's constant with the space-time defects.

All these previous attempts, however, involved very special assumptions which cannot be directly tested by experiments. Also, they cannot give the explicit value of the Planck's constant like what we did in Eq. (37). In our case, the derivation is based strictly on Maxwell's theory, which is well established and is already well supported by experiments.

Recently, quantum communication has become an important new field [30, 31]. Obviously, the physical origin of the Planck's constant has tremendous importance in this area of research. Since *h* essentially defines the size of a bit of information in quantum communication, it is very important to know the physical basis of the Planck's constant. Thus, this work will be helpful to the further development of this field.

## 5. Conclusion

In conclusion, we showed that the energy and momentum of a photon can be determined by treating the photon as a wave packet of electro-magnetic radiation. The following is a summary of the major points of this work:

- The Planck's relation was originally hypothesized based on phenomenological considerations rather than first principles. In order to have a better understanding of this relation, this work shows that it can be derived based on the Maxwell's theory. We found the energy of a photon is indeed proportional to its oscillation frequency.

- The Planck's constant is explicitly given in Eq. (37). We found the value of *h* is closely related to the physical properties of the vacuum.

- The observation of radiation transmitted in discrete energy quanta does not necessarily imply that the photon is a corpuscular object. Instead, the observation of energy quantization only means that the wave packet has a critical size; the wave packet cannot be arbitrarily small. Its emission and transmission follow the principle of all-or-none.

- The "strength" of the wave packet can be characterized by $\zeta$, which represents the integrated strength of the vector potential along a transverse axis. This work suggests that $\zeta$ should have a fixed cut-off value for all photons. Such a cut-off value, in fact, can be determined from the experimental value of the Planck's constant.

- Since *h* is found to be dependent on the *Q* factor of the vacuum, we think it is important to conduct new experiments to accurately measure the linewidth of photons at different frequencies.

**Acknowledgements**

I thank Ms. Lan Fu for her assistance. This work was partially supported by the Research Grant Council of Hong Kong (RGC 660207) and the Macro-Science Program, Hong Kong University of Science and Technology (DCC 00/01.SC01).